\documentclass[pre,amsfonts,twocolumn,amssymb,eqsecnum,floatfix]{revtex4}
\pdfoutput=1
\usepackage{amsmath}
\usepackage{graphicx}
\usepackage{bm}
\usepackage{color}
\allowdisplaybreaks

\begin{document}	
\title{Anomalous diffusion in random-walks with memory-induced relocations}
\author{Axel Mas\'o-Puigdellosas}
\author{Daniel Campos}
\author{Vicen\c{c} M\'endez}
\affiliation{Grup de F{\'i}sica Estad\'{i}stica.  Departament de F{\'i}sica. Facultat de Ci\`{e}ncies. Edifici Cc. Universitat Aut\`{o}noma de Barcelona, 08193 Bellaterra (Barcelona) Spain}

\begin{abstract}
In this minireview we present the main results regarding the transport properties of stochastic movement with relocations to known positions. To do so, we formulate the problem in a general manner to see several cases extensively studied during the last years as particular situations within a framework of random walks with memory. We focus on (i) stochastic motion with resets to its initial position followed by a waiting period, and (ii) diffusive motion with memory-driven relocations to previously visited positions. For both of them we show how the overall transport regime may be actively modified by the details of the relocation mechanism.

\end{abstract}
\maketitle

\section{Introduction}

While Brownian movement is characterized by the well-known diffusive scaling $\langle x^2 \rangle \sim t$, alternative (anomalous) scalings can be obtained when the local motion of the particles is highly non-stationary or somehow governed by heavy-tailed (e.g. power-law) statistics, for instance, in the distribution of the local displacements \cite{MeCaBa14}. The physical mechanisms responsible for such non-standard statistics are wide-ranging, including interaction with the underlying media or with an external force/field, as well as internal mechanisms. In the present mini-review, we focus in the latter case and explore recent advances that have been achieved in exploring anomalous properties of random walk processes when particles are assumed to possess some level of internal memory, such that their displacements are conditioned by the information acquired during its ongoing trajectory. While the framework of random-walks with memory includes many different models (as self-avoiding walks \cite{MaSl96}, elephant or alzheimer random-walks \cite{ScTr04,CrVi12}, infotactic strategies \cite{VeVi07},...) we focus here in the particular case where particles use their memory from time to time to relocate to known, or familiar, positions in the domain. 

As a first case of interest, relocations to the initial position (resets) have been recurrently studied in recent years. One-dimensional unbounded diffusion with resets happening at a constant rate was initially introduced in \cite{EvMa11}. Originally its interest was focused on their ability to make the mean first passage time finite, with a minimum value found for an intermediate reset rate \cite{EvMa11p,PaRe17,ChSo18,PaKuRe19}. However, it has been subsequently shown that their intrinsic transport properties are also of interest. For example, as a consequence of such resetting, dispersal is asymptotically suppressed and a steady state is reached. Thereafter, this property has been confirmed by numerous works on Markovian resets in different contexts, as multi-dimensional diffusion \cite{EvMa14}, coagulation-diffusion processes \cite{DuHe14}, confined diffusion \cite{Pa15,ChSc15}, diffusion with a refractory period after the resets \cite{EvMa19}, anomalous subdiffusion \cite{Sh17,KuGu19}, monotonic stochastic motion \cite{MoVi13,MoMa17}, continuous-time random walk (CTRW) velocity models \cite{MeCa16}, the telegraphic process \cite{Ma19} and underdamped Brownian motion \cite{Gu19}. Likewise, in \cite{PaKu16}, a steady state is shown to appear when a diffusion process is restarted at a time-dependent rate and in \cite{NaGu16} power-law reset time probability density functions (pdf) are considered and conditions for a steady state to exist are found. Finally, general conditions on the reset time pdf for the appearance of a steady state have been found in \cite{EuMe16,MaCaMe19}.

In contrast with the aforementioned cases, where a steady state is reached, some works have shown that unbounded dispersal is still possible at a population level when the reset time is governed by heavy-tailed distributions so making resets less and less frequent with time (see \cite{MeCa16,EuMe16,MaCaMe19,MaCaMe19_1}). Particularly, in \cite{MaCaMe19_1,EvMa19} it is shown that the diffusivity of a walker which resets its position followed by a residence or refractory period at the origin is strongly dependent on the tail of the reset and residence time pdfs. Also, asymptotic transport appears when the resetting is soft \cite{KuGu19}, meaning that the walker is relocated to the origin but the other properties of the motion (e.g. a dynamic diffusion coefficient) are not renewed.

Alternatively, random walks with relocations to any previously visited place have been seen to modify the transport regime of a given motion process. In \cite{BoSo14} it was proved that if such relocations are equiprobable among all visited sites in the past, then unbounded dispersal is not suppressed as in the resetting case but becomes ultraslow, with a mean squared displacement (MSD) that grows logarithmically (i.e. $\langle x^2(t) \rangle \sim \ln(t)$), a result that is kept when relocation events follow a time-continuous dynamics \cite{BoSo14,BoEv17,CaMe19}. This can be generalised to include a weight function for the memory to chose the relocation position; in such case, the MSD of the process can exhibit a range of behaviors, from diffusive or sub-diffusive to logarithmic, as a function of the distribution \cite{BoRo14}. Also, a rich variety of transport regimes have been proved to arise in the continuous time and space version of this model \cite{BoEv17}, ranging from an ultraslow growth $\langle x^2(t) \rangle \sim \ln(\ln(t))$ to the diffusive scaling. Finally, a more specific relocation mechanism consisting of stochastically taking the walker to the maximum position attained in the past has also been proven to also let the motion spread \cite{MaSa15}.

In the following, we employ a general framework that includes these two types of models (resets and/or relocations to visited sites) as particular cases, and so allows us to review the results mentioned above from the unified perspective of random-walks with memory-induced jumps, and detect opportunities of research in the field for the near future. We focus our efforts on recovering the different transport regimes mentioned above, so illustrating the capacity of internal memory to modify the scaling properties governing the transport regime of the process.

\section{General Framework}
\label{SecFramework}

The framework we consider here follows a time dynamics based on the alternation between two states (one for standard motion and another for relocations), an approach which is quite usual in models of random walks with memory \cite{KuHa18,MaCaMe19_1}. First, a normal state ($i=1$) where the walker motion is governed by a given by a jump length distribution and a probability time distribution, as in the classical CTRW. The duration of this state is determined by a given pdf $\varphi_1(t)$. Second, a memory-induced state ($i=2$) which results from introducing a relocation to a particular position without explicit dependence of memory and waiting there until a new normal period is started. The relocation position is chosen from a generalized relocation distribution $p_0(x,t)$. After the relocation takes place, we assume that there exists a \textit{refractory}, or waiting, period during which the particle remains at the position of relocation, governed by another pdf denoted as $\varphi_2(t)$.

If the walker starts at $t=0$ from position $x=0$ at state $i=1$, the transition probability $j_1(x,t)$ from state $i=2$ to $i=1$ at position $x$ and the transition probability $j_2(x,t)$ from state $i=1$ to $i=2$ at $x$ will follow, respectively:

\begin{align}
j_1(x,t)&=\delta(x)\delta(t)+\int_0^t j_{2}(x,t-t')\varphi_{2}(t')dt'
\label{Eq1}
\\
j_2(x,t)&=p_0(x,t)\int_{-\infty}^{+\infty}dx'\int_0^t j_{1}(x',t-t')\varphi_{1}(t')dt'.
\label{Eq2}
\end{align}\\

Let us now introduce the spatial dynamics for both states. In the normal state, the motion can be described by a general propagator $P(x,t;x',t')$, being the probability of finding the walker at point $x$ at time $t$ if it was at point $x'$ at time $t'$. Otherwise, in the memory-induced state the walker stays at the relocation position $x'$, so its 'propagator' reduces to $\delta(x-x')$. As a whole, the pdf of the particles in state $i=1$ and $i=2$ at time $t$ respectively read

\begin{align}
\rho_1(x,t)&=\int_{-\infty}^{+\infty}dx'\int_0^t dt' j_1(x',t-t')\varphi_1^*(t')P(x,t';x',0)
\label{Eq3}
\\
\rho_2(x,t)&=\int_{-\infty}^{+\infty}dx' \int_0^t dt' j_2(x', t-t')\varphi_2^*(t') \delta(x-x'),
\label{Eq4}
\end{align}
where $\varphi_{i}^*(t) \equiv \int_t^{\infty}\varphi_{i}(t')dt'$, for $i=\{ 1,2 \}$. The meaning of the first equation can be stated as follows: the pdf for particles in the memory-free state ($i=1$) is described by the propagator $P(x,t';x',0)$, provided the system entered this state at time $t-t'$ at any position $x'$, and it has remained in that state (i.e. without relocating) for the subsequent time $t'$. Eq. \eqref{Eq4} represents the equivalent for the memory-induced state, with the position described by the delta function $\delta(x-x')$ instead of the propagator.

\section{Spatial dispersal with resets}
\label{SecResets}

Resets can be defined as relocations which are used by the particle to come back from time to time to its initial position, an idea which can be satisfactorily adapted to study situations like animal foraging \cite{CaMe15,PaKuRe19}, searches on the Internet \cite{ViVi91,ToFaPa08} or genetic networks \cite{KoHoRo08,LiWaAnKoHu17}, and the kinetics of chemical reactions \cite{ReUrKl14,Reuveni16} or molecular proofreading \cite{BaTlLi02,MuHuLe12}. Using the general formalism in the previous section, this corresponds to a time-independent relocation distribution $p_0(x,t)=\delta(x)$. Also, for the sake of simplicity we can restrict our analysis to propagators which are space homogenous, such that motion in the memory-free state satisfies $P(x,t;x',0)=P(x-x',t;0,0)\equiv P(x-x',t)$. 

Performing the Fourier-Laplace transform of the equations (\ref{Eq1}-\ref{Eq4}) above and solving for the MSD of the overall propagator as $\langle x^2(t)\rangle \equiv \int_{-\infty}^{\infty} dx\ x^2 \rho(x,t) = \int_{-\infty}^{\infty} dx\ x^2 \left( \rho_1(x,t)+\rho_2(x,t) \right)$, it can be found that

\begin{equation}
\mathcal{L} \left[ \langle x^2(t)\rangle \right] = \frac{\mathcal{L}\left[\varphi_1^*(t)\langle
x^2(t)\rangle_P \right]}{1-\hat{\varphi}_1(s)\hat{\varphi}_2(s)},
\label{Eq6}
\end{equation}
where $\mathcal{L}[f(t)] = \hat{f}(s) \equiv \int_0^\infty e^{-st}f(t) dt$ represents the Laplace transform of $f(t)$, and $\langle x^2(t)\rangle_P$ is the MSD of the propagator $P(x,t;x_0 ,0)$ by assuming that it is symmetric in space.

Seminal works on random walks with resets were focused on the situation where the process restarts immediately after the reset happens, i.e. $\varphi_2(t)=\delta(t)$ \cite{EvMa11,EvMa11p}. In such case, it is known that if the memory-free propagation scales as $\langle x^2(t)\rangle_P\sim t^p$ and the resets are Markovian (i.e. $\varphi_1(t)=r e^{-r t}$), a stationary state is always reached (so, there is propagation failure), and the relaxation to that stationary situation can be also characterized \cite{MaSaSc15}. This scenario, where the resetting is strong enough to localize the walker around the origin, has been later studied from different perspectives. In some cases, a modified Fokker-Planck equation formalism \cite{EvMa11,EvMa14,DuHe14,Pa15,ChSc15,Gu19,Ma19} has been employed to study Markovian resetting in, for instance, the diffusion equation \cite{EvMa11}, the Telegrapher's equation \cite{Ma19} and the underdamped Brownian motion equation \cite{Gu19}. Other works have found this same result by interpreting resets as a renewal of the motion and consequently building a renewal master equation for the overall pdf as done in the general formalism herein \cite{Sh17,KuGu19,MoVi13,MoMa17,MeCa16}. 

Otherwise, for the same propagation scaling but a long-tailed statistics of reset times with diverging first moment ($\varphi_1(t)\sim t^{-1-\gamma_1}$, with $0<\gamma_1<1$), the overall MSD scales also as $\langle x^2(t) \rangle\sim t^p$. Therefore,  the properties of the memory-free movement can be modified or not depending on the specific properties of the refractory or waiting time. This was found in \cite{EuMe16} for diffusive motion (i.e. $p=1$) and later in \cite{MaCaMe19} for the general case.

The model turns out to be more interesting, however, when the resting time after the resets is non-zero, this is, for a non-trivial choice of the waiting time pdf as $\varphi_2(t) \sim t^{-1-\gamma_2}$. This scenario, which has been recently studied in \cite{EvMa19,MaCaMe19_1,PaKuRe19} from a renewal perspective, yields a wide range of situations for the asymptotic behavior of the overall MSD in the case when memory-free propagation scales again as $\langle x^2(t) \rangle_P\sim t^p$. Depending on the finiteness/infiniteness of the moments of $\varphi_1(t)$ and $\varphi_2(t)$, the following cases can be identified:

\begin{itemize}
\item[i)] $\gamma_1>1$, $\gamma_2>1$. The propagation ceases and a stationary state is reached as for the non-resting period case, yielding 
\begin{equation*}
\langle x^2(t)\rangle\sim \text{const.}
\end{equation*}
\item[ii)] $\gamma_1>1$, $0<\gamma_2<1$. The propagation ceases and the system tends to collapse towards the origin.
\begin{equation*}
\langle x^2(t)\rangle\sim t^{\gamma_2-1}
\end{equation*}
\item[iii)] $0<\gamma_1<1$, $\gamma_2>1$. The propagation is only affected by the resetting mechanism in a multiplicative factor, but it does not affect the scaling:
\begin{equation*}
\langle x^2(t)\rangle\sim t^{p}
\end{equation*}
\item[iv)] $0<\gamma_1<1$, $0<\gamma_2<1$. The propagation is actively modified by the resetting mechanism when the tail of $\varphi_2\sim t^{-1-\gamma_2}$ is longer than the tail of $\varphi_1\sim t^{-1-\gamma_1}$, i.e. $\gamma_2<\gamma_1$. Otherwise, when $\gamma_2\geq \gamma_1$, the overall process behaves as in case iii). In short,
\begin{equation*}
\langle x^2(t)\rangle\sim t^{p-(\gamma_1-\gamma_2)\theta(\gamma_1-\gamma_2)},
\end{equation*}
where $\theta(\gamma_1-\gamma_2)$ denotes the Heaviside function.

\end{itemize} 

\begin{figure}
\includegraphics[scale=0.5]{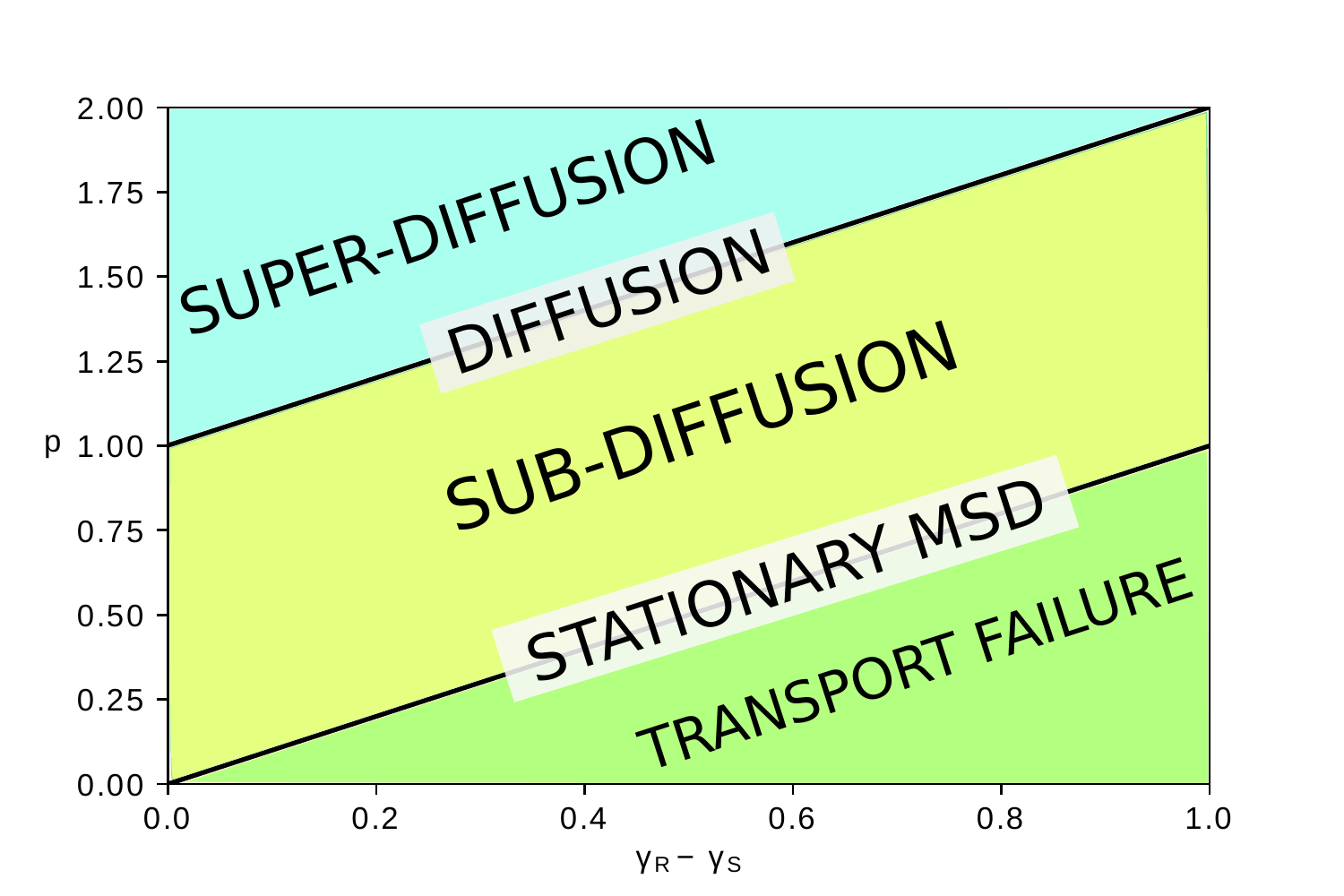}
\caption{Schematic plot of the different overall transport regimes in terms of the transport regime of the movement process $p$ and the difference between the decaying exponents of the reset and residence time pdfs $\gamma_1-\gamma_2$, corresponding to case iv) in the main text. In the blue region the overall behaviour is super-diffusive. In the yellow region it is sub-diffusive and the green represents the values for which we have transport failure. We can see that, for instance, an originally super-diffusive process ($p>1$) can become sub-diffusive (yellow region) with the properly chosen tail exponents.}
\label{Fig1}
\end{figure}

According to this, scenario iv) turns out to be particularly interesting and all its casuistic is visually summarized in Figure \ref{Fig1}. When the resting period is asymptotically longer than the active period, the diffusivity of the propagation is reduced by a factor $\gamma_1-\gamma_2$ as a result of the competition between the heavy-tailed effects of the movement and the refractory period. This, for instance, may turn a superdiffusive (or diffusive) process into subdiffusive by only tuning the asymptotic decay of the active and resting times pdf. 

In the light of what we have seen in this section, some questions regarding the asymptotic transport properties of motion with resetting remain still unanswered. For instance, without the residence time after the resetting, it seems that resetting either leaves the transport regim unaltered or it makes the transport cease and a stationary state is reached. Is there any resetting mechanism able to smoothly modify the transport regime of the motion? Also, resetting has been mainly treated as an internal mechanism of the motion. What would be the overall dynamics of a set of walkers which interact to suddenly reset their individual position? Despite this has been shown to be hard from an analytical point of view \cite{FaEv17}, it would be extremely interesting for the description of many ecological systems.

\section{Spatial dispersal with relocations to visited places}
\label{Relocations}

Relocation processes in which particles are allowed to return to already visited places is another case of interest for which the properties of the overall MSD have been recurrently explored \cite{BoRo14,BoSo14,CaMe19,FaBoGiMa17}. If expressed in terms of the general framework presented in Section II, this case would correspond to a relocation distribution of the type

\begin{equation}
p_0(x,t)=\frac{\int_0^t \phi(t')\rho(x,t')}{\int_0^t \phi(t')dt'},
\end{equation}
where $\phi(t)$ is a memory function which weights all possible relocation places as a function of the time elapsed since they were visited. This is, if $\phi(t)$ is an increasing function, then relocation to recent positions is more likely to occur, which would implicitly assume that memory is a vanishing process. On the other hand, if $\phi(t)$ is a decreasing function of time then the initial positions are the most probable ones. Finally, one could even recover the resetting mechanism by choosing $\phi(t) \sim \delta(t)$. 

Since the mathematical treatment for a general situation as a function of $\phi(t)$ becomes cumbersome, we can restrict ourselves to Markovian relocations (so $\varphi_1(t)= r e^{-rt}$) and nule resting times after the relocation, $\varphi_2(t)=\delta(t)$. For this particularly simple case, the four general Eqs. (\ref{Eq1}-\ref{Eq4}) lead to the following implicit solution for the overall MSD; if the propagation is diffusive with $\langle x^2(t)\rangle_P= 2Dt$, we get

\begin{equation}
\mathcal{L} \left[ x^2(t)\right]=\frac{2 D}{s (r+s)}+\frac{r}{r+s} \mathcal{L}\left[ \frac{\int_0^t \phi(t')\langle x^2(t')\rangle}{\int_0^t \phi(t')dt'}\right].
\end{equation}

From this expression, the long time behaviour of the MSD can be deduced for different weight functions for the memory; this has been done, using a relatively different perspective, in \citep{BoEv17}. There it is shown that up to five different regimes emerge, which illustrates the rich variety of the model:

\begin{itemize}
\item[i)] $\phi(t)\sim t^{-a}$, with $a>1$.
The transport ceases and a stationary state is reached. 
\begin{equation*}
\langle x^2(t)\rangle \sim \text{const.}
\end{equation*}
\item[ii)] $\phi(t)\sim t^{-a}$, with $a=1$.
The transport becomes extremely slow.
\begin{equation*}
\langle x^2(t)\rangle \sim \ln (\ln (t))
\end{equation*}
\item[iii)] $\phi(t)\sim t^{-a}$, with $a<1$.
The transport becomes ultra slow.
\begin{equation*}
\langle x^2(t)\rangle \sim \ln(t)
\end{equation*}
\item[iv)] $\phi(t)\sim e^{t^b}$, with $0<b\leq 1$.
The transport becomes sub-diffusive when $b<1$ and it is not affected by the memory for $b=1$.
\begin{equation*}
\langle x^2(t)\rangle \sim t^b
\end{equation*}
\item[v)] $\phi(t)\sim e^{t^b}$, with $1<b$.
The transport is not qualitatively affected by the memory, so it remains diffusive.
\begin{equation*}
\langle x^2(t)\rangle \sim t.
\end{equation*}
\end{itemize}

These results have progressively been found during the last years. While case iii) was originally found in \cite{BoSo14} and later in \cite{CaMe19}, scalings iv) and iv) were found in \cite{BoRo14} and, finally, all the asymptotic scalings derived herein were found in \cite{BoEv17}. Likewise, transitions between different transport regimes have can possibly emerge as a consequence of spatial \cite{BoFaGiMa19} or temporal \cite{FaBoGiMa17} heterogeneities in the resetting process.

This range of situations is already obtained for fixed (and particularly simple) forms of $\varphi_1(t)$ and $\varphi_2(t)$, while the situation is susceptible to become even more complex as more general pdfs are taken into account. All this enlights the theoretical interest of the memory-induced mechanism as a way to reproduce different propagation regimes.

\section{Future perspectives}
\label{Summary}

Mostly inspired by the movement of biological individuals and other systems of \textit{intelligent} walkers, random-walk models with memory-induced relocations (either resets or relocation to visited places) have been widely explored in recent years. In this minireview we have tried to condense the current knowledge we have about their transport properties in order to illustrate the richness of macroscopic transport properties they are able to yield. New situations of interest can arise in the future as long as different choices for the relocation distribution $p_0(x,t)$ in Eq.(\ref{Eq2}) are explored instead of the two (resetting, uniform relocation) reviewed here. This may include the case of resetting to a distribution of fixed points, visited or not (a topic which has also received attention in the biological literature \cite{Gautestad11,Gautestad15}), or a relocation dynamics based on returns to those sites that were more beneficial in the past (so introducing an additional variable representing food available or assigning a value to the visited sites). As long as more general models and conclusions are obtained, a meaningful comparison to real data from animals or other organisms can represent a promising way to explore memory capacities in these living systems; or, alternatively, they can also become a useful tool to study the properties of active matter when subject to memory effects \cite{FaBoRa19}. So, we envision that the following years will probably witness an increasing interest of researchers for the intricate interplays between memory and transport properties.

\section*{Acknowledgements}
This research has been supported by the Spanish government through Grant No. CGL2016-78156-C2-2-R.

\bibliography{Referencesa}

\end{document}